\documentstyle[11pt,]{article}
\begin{document}
\centerline{\bf{GRAVITATIONAL WAVES IN}}
\centerline{\bf{GENERALISED BRANS-DICKE THEORY}}
\vspace{0.2in}
\centerline{B.K.Sahoo$^{1}$ and L.P.Singh$^{2}$}
\centerline{Department of Physics,}
\centerline{Utkal University,Bhubaneswar-751004,India}
\centerline{$^{1}$ bijaya@iopb.res.in}
\centerline{$^{2}$ lambodar@iopb.res.in}
\vspace{0.25in}
\begin{abstract}
We have solved Cosmological Gravitational Wave(GW) equation in the frame
work of Generalised Brans-Dicke(GBD) theory for all epochs of the
Universe.The solutions are expresed in terms of the present value of
the Brans-Dicke coupling parameter $\omega(\phi)$.It is seen that the
solutions represent travelling growing modes for negative values of
$\omega_{0}$ for all epochs of the Universe.
\end{abstract}
Key words:Gravitational waves,growing modes.\\
PACS NO:98.80.-K,98.80.cq
\section{INTRODUCTION}
  General Relativity (GR) predicts that perturbations of a background
  spacetime  generate gravity waves.The general results concerning
  gravitational  waves have been investigated by numerous authors 
long ago[1].In the context of GR cosmologies,Grishchuk [2] has shown 
that the 
  varying gravitational field of the expanding Universe would amplify
  zero -point fluctuations, and lead to the formation of relic
  gravitons .Gravitons interact very weakly with matter and other
  fields, and only at energies of order of the planck
  enegy($~10^{19}$GeV) will the gravitational interaction become as
  strong as the electromagnetetic interaction.Therefore,the relic
  gravitons must have decoupled from the dynamics of the Universe at
  extremely early times,and offer a unique way to probe the
  structure and evolution of the early Universe.Even without detecting 
  them,we can still get useful information by studying the
  consequences of their presence in several processes of the cosmic
  history ;for example,primordial nucleosynthesis and the anisotropy
  they would induce in the microwave background .Thus,the idea of
  using gravitational radiation as probe of the  early Universe has
  turned into a realistic possibility.
            The subject of gravitational radiation within Brans-Dicke
  (BD) theory was carried out by Wagoner[3].This was concerned with 
  gravitational waves generated in the weak field limit.Barrow[4]has
  derived the cosmological gravitational wave equation in  some more
  general scalar tensor theories of gravity and solved the wave
  equation for universe in BD theory where the coupling function is 
   a constant.However,in the generalised
  Brans-Dicke(GBD) theory of gravity  where the coupling function
  $\omega=\omega(\phi)$,the gravitational wave  equation is solved by Barrow for only
  vacuum and radiation dominated epochs of the Universe for a
  particular choice of the coupling function $\omega(\phi)$.Thus the
  work seems to be incomplete in GBD theory of gravity.Here we
  investigate  the cosmological solutions of the gravitational wave  equation
  for all epochs of the Universe in GBD theory  by allowing the
  coupling parameter to run with scalar field $\phi$ following the
  approach adopted in our recent work[5].The solutions are 
  expresed in terms of present value of the coupling parameter
  $\omega(\phi)$ i.e,in terms of $\omega_{0}$.
             An outline of this paper is as follows.In the section II
             we have derived the solutions.The solutions are expresed
             in terms of $\omega_{0}$ in section III for all epochs of 
             the Universe and shown that for negative values of
             $\omega_{0}$ solutions represent growing modes..We conclude with section IV.
\section{WAVE EQUATION }
The gravitational field equations are given below[6]

 \noindent$R_{\mu \nu}=\frac{8\pi G}{\phi}[T_{\mu
   \nu}-\frac{1+\omega}{2\omega+3} g_{\mu \nu}
     T]+\frac{\omega}{\phi^{2}} \phi,_{\mu} \phi,_{\nu}+\frac{1}{\phi}
   \phi,_{\mu \nu} 
-\frac{1}{2}\frac{g_{\mu\nu}}{\phi}
\frac{d\omega/d\phi}{2\omega+3} \phi_{,\sigma}
\phi^{,\sigma}-g_{\mu\nu} \frac{1}{2\omega+3}\left[\phi(dU/d\phi)+2(\omega+1)U\right]$

\begin{equation}
\Box\phi =\frac{8\pi}{2\omega+3}T-\frac{d\omega/d\phi}{2\omega+3}
      \phi_{,\sigma} \phi_{,\sigma} -\frac{2}{2\omega+3}
      [\phi^{2}(dU/d\phi)-\phi U]
\end{equation}
When perturbation is applied to above equations,they become[4,7]\\

\noindent $\delta R_{\mu \nu}=\delta( \frac{8\pi G}{\phi}[T_{\mu
  \nu}-\frac{1+\omega}{2\omega+3} g_{\mu \nu}
T])+\delta(\frac{\omega}{\phi^{2}} \phi_{,\mu}
\phi_{,\nu})+\delta(\frac{1}{\phi} \phi_{,\mu \nu})-\delta(\frac{1}{2} 
\frac{g_{\mu \nu}}{\phi} \frac{1}{2\omega+3} \frac{d\omega}{d\phi}
  \phi_{,\sigma} \phi^{,\sigma})-\delta(g_{\mu \nu}
  \frac{1}{2\omega+3} [\phi \frac{dU}{d\phi} + 2(\omega + 1)U])$\\

$\delta \Box \phi = \delta(\frac{8\pi T}{2\omega+3})-
\delta(\frac{1}{2\omega+3} \frac{d\omega}{d\phi} \phi_{,\sigma}
\phi_{,}^{\sigma})-\delta(\frac{2}{2\omega+3}
[\phi^{2}\frac{dU}{d\phi}-\phi U])$\\

\noindent Where the perturbed metric is\\
$\tilde{g}_{\mu \nu}=g_{\mu \nu}+\delta g_{\mu \nu}$ \\
\hspace{.5in}$=g_{\mu \nu}+h_{\mu \nu},$ \hspace{1in}where,$ \delta
           g_{\mu\nu}=h_{\mu \nu}$ \\
\noindent leading to the Ricci tensor, $\tilde{R}_{00}=R_{00}+\delta R_{00}$.\\
The energy-momentum tensor is given by 
, $\tilde{T}^{00}=T^{00}+\delta T^{00}=T^{00}+\delta \rho,$\\
\noindent The trace of  energy-momentum tensor  is 
, $\tilde{T}=T+\delta T=T+\delta \rho-3 \delta P$.\\
\noindent For the d'Alembertian of the scalar field we find
, $\delta \Box  \phi=\delta \ddot{\phi}+a \dot{a} h^{ij} \dot{\phi}-\frac{1}{2a^{2}} \dot{h_{ij}}\dot{\phi}+3\frac{\dot{a}}{a} \delta\dot{\phi}-\frac{\nabla^{2}}{a^{2}} \delta\phi$.\\
\noindent We adopt the synchronous gauge, $h_{0\mu}=0$\\
\noindent The time time component of perturbed Ricci tensor is[8], 
$ \delta R_{00}=\frac{1}{a^{2}}\left[ \ddot{h_{ij}}-2\frac{\dot{a}}{a} \dot{h_{ij}}+2(\frac{\dot{a}^{2}}{a^{2}}-\frac{\ddot{a}}{a})h_{ij}\right].$\\
\noindent When $\phi$=constant,all the perturbed equations will reduce to perturbed equation of Eienstein.
The perturbations to the metric that represent weak gravitational waves can be expressed as[9] \\
\centerline{$h_{ij}(t,x)=\int d^{3}k\hspace{.1in} h_{ij}^{(k)}(t,x),$}\\
with,
\begin{equation}
h_{ij}^{(k)}(t,x)=\frac{1}{a^{2}(t)}Y_{k}(t)\hspace{.1in} \zeta _{ij}(k,x),
\end{equation}
Where the functions $Y_{k}(t)$ and $\zeta_{ij}(k,x)$satisfy\\
\centerline{$(\nabla^{2}+k^{2})\hspace{.1in}\zeta_{ij}(k,x)=0$,}
\begin{equation}
\ddot{Y_{k}}+f(t) \dot{Y_{k}}+g(t)Y_{k}=0,\\
\end{equation}
where \hspace{1in}$ f(t)=\frac{\dot{\phi}}{\phi}-\frac{\dot{a}}{a},$\\
$g(t)=\frac{k^{2}}{a^{2}}+4\frac{\dot{a}^{2}}{a^{2}}-\frac{8\pi}{\phi}\left[\frac{2(1+
\omega)}{2\omega+3} \rho - \frac{2\omega}{2\omega+3} P\right]
 -\frac{1}{2\omega+3} \frac{d\omega}{d\phi} \frac{\dot{\phi}^{2}}{\phi}+\frac{1}
{2\omega+3}\left[\phi \frac{dU}{d\phi}+2(\omega+1)U\right]$\\
In the above equations,a(t) is the scale factor, $i,j = 1,2,3;$  $\nabla^{2}$ 
is the spatial Laplacian operator, an over dot indicates derivatives with 
respect to the synchronous cosmic time ,t, and $ k=
|\vec{k}|=\frac{2\pi a}{\lambda}$ with $\vec{k}$ 
being the comoving wave vector.The eigen functions $\zeta_{ij}(k,x)$  can be written in terms of  plane -wave 
solutions $exp(\pm i \vec{k}. \vec{x})$ times a constant polarization tensor  
and using the field equations in the zero curvature FRW model we can simplify 
(3) and recast it as [4]
\begin{equation}
\ddot{Y_{k}}+\left[\frac{\dot{\phi}}{\phi}-\frac{\dot{a}}{a}\right]\dot{Y_{k}}+\left[\frac{k^{2}}{a^{2}}-
2\frac{\ddot{a}}{a}-2\frac{\dot{a}}{a} \frac{\dot{\phi}}{\phi}-\frac{d\omega
/d\phi}{2\omega+3} \frac{\dot{\phi}^{2}}{\phi}+\frac{1}{2\omega+3} (\phi \frac{dU}{d\phi}+2(\omega+1)U)\right]Y_{k}=0
\end{equation}
This is the Gravitational wave equation in more general scalar-tensor theory of gravity.
\section{SOLTUIONS IN GBD THEORY }
The generalised Brans-Dicke scalar-tensor theory of gravity is
characterised by the generalisation of the coupling function $\omega$
to  a function of $\phi$; and $U=\frac{dU}{d\phi} =0$.We introduce the
conformal time $\eta$ defined by $dt=a d\eta$\\  Each component $h_{ij}^{(k)}$ of the gravitational wave perturbations can be written as[
4] 
\begin{equation}
h_{ij}^{(k)}(\eta,x)=\frac{1}{R(\eta)} \mu(k,\eta) \zeta_{ij}(k,x)
\end{equation}
Here $ R(\eta)=a(\eta)\phi^{1/2}(\eta)$ and 
\begin{equation}
{\mu^{\prime \prime}}(k,\eta)+\left[k^{2}-\frac{{R}^{\prime \prime}}{R}-\frac{d\omega/d\phi}{2\omega+3} \frac{{\phi ^{\prime}}^{2}}{\phi}\right]\mu(k,\eta)=0
\end{equation}
with the prime indicating derivatives with respect to $\eta$.\\
\noindent When $\omega(\phi)$= constant, eqn.(6) becomes, 
$\mu^{\prime \prime}(k,\eta)+\left[k^{2}-\frac{R^{\prime \prime}}{R}\right]\mu(k,\eta)=0$\\
\noindent which represents Gravitational wave equation in Brans-Dicke scalar-tensor theory
of gravity.The solution given by eqn(6)  is formally identical to the corresponding one obtained in the general relativistic case,but with the quantity $R(\eta)$ defined by eqn(5) replacing the scale factor $a(\eta)$.Equation (6) describes an oscillator with varying frequency or a Schrodinger equation with potential\\
$\frac{R^{\prime \prime}}{R}+\frac{d\omega/ d\phi}{2\omega+3} \frac{{{\phi^{\prime}}}^{2}}{\phi}$\\
for a particle with energy $k^{2}$ and with the variable $\eta$ playing the role of a spatial co-ordinate.
We solve eqn(6) for different epochs of Universe,with equation of
state $P=\gamma \rho$,  $-1\leq \gamma \leq 1$.\\
The scale factor  and the scalar field evolve as power laws of the cosmic time
\begin{equation}
a(t)=a_{0}\left(\frac{t}{t_{0}}\right)^{\alpha}
\end{equation}
\begin{equation}
\phi(t)=\phi_{0}\left(\frac{t}{t_{0}}\right)^{\beta}
\end{equation}
\centerline{$\omega(t)=\omega(\phi) \approx
  \phi^{n}=\omega_{0}\left(\frac{t}{t_{0}}\right)^{n\beta}$}.\\
For $\alpha \neq 1$,the translation of eqn(7) and (8) into the
conformal time yields\\
$a(\eta)=C_{0}\left(\eta-\overline{\eta_{0}}\right)^{q}$\\
$\phi(\eta)=\sigma_{0}\left(\eta-\overline{\eta_{0}}\right)^{r}$\\
$R(\eta)=c_{0}\left(\eta-\overline{\eta_{0}}\right)^{s}$\\
where
$q=\frac{\alpha}{1-\alpha},r=\frac{\beta}{1-\alpha},s=q+\frac{r}{2}=\frac{\alpha}{2(1-\alpha)}$
and \\
$\overline{\eta_{0}} \equiv \eta_{0}-\frac{q}{a_{0} H_{0}},
\eta_{0}\equiv \eta(t_{0}),H_{0}\equiv H(t_{0}),c_{0}\equiv 
a_{0}\left( \frac{a_{0}H_{0}}{q}\right)^{q},\sigma_{0}\equiv \phi_{0}
\left( \frac{a_{0}H_{0}}{q}\right)^{r},\\ C_{0}\equiv
  c_{0}\sigma_{0}^{1/2}=R_{0} \left( \frac{a_{0}
      H_{0}}{q}\right)^{s},R_{0}\equiv a_{0} \phi_{0}^{1/2}.$\\

\noindent The wave equation (6)takes the form\\
\begin{equation}
\mu^{\prime \prime}(k,\eta)+\left[k^{2}-\frac{s(s-1)}{(\eta-\eta_{0})^{2}}-\frac{d\omega/d\phi}{2\omega+3}    \frac{{\phi^{\prime}}^{2}}{\phi}\right]\mu(k,\eta)
\end{equation}
We take $2\omega+3=(2\omega_{0}+3)\phi^{n}$\\
This leads to $2 \frac{d\omega}{d\phi}=(2\omega_{0}+3)n \phi^{n-1}$.\\
This reduces  equation (9)to\\

$\mu^{\prime \prime}+\left[k^{2}-\frac{1}{(\eta-\overline{\eta_{0}})^{2}}\left(s(s-1)-\frac{nr^{2}}{2}\right)\right]\mu=0$\\

\noindent Let $s(s-1)-\frac{nr^{2}}{2}=s^{\prime}(s^{\prime}-1)$\\
Where $s^{\prime}=\frac{1}{2} \pm \frac{1}{2} \sqrt{4s^{2}-4s+1-2nr^{2}}$\\

\noindent Hence,we get  
\begin{equation}
\mu^{\prime \prime}+\left[k^{2}-\frac{s^{\prime}(s^{\prime}-1)}{(\eta-\overline{\eta_{0}})^{2}}\right]\mu=0
\end{equation}
For very large wave numbers or,more precisely, for waves such that\\

$k^{2}\gg V(\eta),$
where $V(\eta)\equiv|R^{\prime \prime}/R
+\frac{d\omega/d\phi}{2\omega+3}\frac{{\phi^{\prime}}^{2}}{\phi}|\equiv
|\frac{s^{\prime}(s^{\prime}-1)}{(\eta-\overline{\eta_{0}})^{2}}|$,\\

\noindent  eqn(10) will have purely oscillatory solutions, and eqn(5) takes the form \\

$h_{ij}^{(k)}(\eta,x)=\frac{1}{a(\eta)\phi^{1/2}(\eta)} \left[
    C_{1}\epsilon^{ik(\eta-\overline{\eta_{0}})} +C_{2}
    \epsilon^{-ik(\eta-\overline{\eta_{0}})}\right]\zeta_{ij}(k,x)$\\
where $C_{1}$ and $ C_{2}$ are arbitrary constants and the third constant 
$\overline{\eta_{0}}$ is same as before.We see that the effect of the 
scalar field $\phi$ is for the scale of amplitude decrease to be set
by the function $R(\eta)$ instead of by $a(\eta)$.\\

\noindent In the opposite regime of very small wave numbers, where $k^{2}\ll
V(\eta)$,\\

\noindent we obtain,\hspace{.5in} $h_{ij}^{(k)}(\eta,x)=\left[ C_{1}+C_{2}
  \int{}^{\eta} R^{-2}(\eta ^{\prime})d\eta^{\prime}\right]
\zeta_{ij}(k,x)$\\

\noindent Note ,however, that ,for bell-shaped potentials, the amplification
coefficient will scale as $R(\eta_{f})/R(\eta_{i})$ instead of $a(\eta_{f})/a(\eta_{i})$.\\
The general solution of eqn(10) can be written as[10]
\begin{equation}
\mu(k,\eta)=k^ {-\frac{1}{2}} X^{\frac{1}{2}}\left[C_{1} J_{m}(X) +C_{2}Y_{m}(X)\right]
\end{equation}
Here $J_{m}$ and $Y_{m}$ are Bessel  and Neumann's functions  respectively ,$X(\eta)=k(\eta-\overline{\eta_{0}})$ and \\
\noindent $m=s^{\prime}-1/2$\\
\hspace{.1in}$=\pm\frac{1}{2}\sqrt{4s^{2}-4s+1-2nr^{2}}$\\
$=\pm\frac{1}{2}\sqrt{ 4\left(\frac{2\alpha+\beta}{2(1-\alpha)}\right)^{2}-
4\left(\frac{2\alpha+\beta}{2(1-\alpha)}\right) +1 -2n \left( \frac{\beta}{1-\alpha}\right)^{2}}$\\
So,\[\mu=(\eta-\overline{\eta_{0}})^{1/2}\left[ C_{1}\sum_{r=0}^{\infty} A_{r} \left(\frac{X}{2}\right)^{m+2r} + C_{2} \sum_{r=0}^{\infty} B_{r} \left(\frac{X}{2}\right)^{-m+2r} \right]\]
\[=(\eta-\overline{\eta_{0}})^{1/2}\left[ C_{1} \sum_{r=0}^{\infty} A_{r} \left(\frac{\pi a(\eta)}{\lambda}(\eta-\overline{\eta_{0}})\right)^{m+2r}  + C_{2}\sum_{r=0}^{\infty} B_{r}\left( \frac{\pi a(\eta)}{\lambda}(\eta-\overline{\eta_{0}})\right)^{-m+2r}\right] \]
We see that in order to represent $\mu$ a physical mode the power $m$ has to be real and greater than zero.
The difference between the solution (11) and the corresponding one in
GR lies in the presence of the coefficient $\beta$ in the
parameters $s^{\prime}$  and m.Setting  $\beta=0$  we recover the GR
solution as  expected, since this corresponds to the requirement that
$\phi$ ,and hence G,  remains constant.
The general cosmological solutions in the framework of GBD theory when
matter satisfies the equation of state $P=\gamma \rho$ with $-1\leq
\gamma  \leq 1$ a constant,were derived by Sahoo and Singh [11].An
important feature of these solutions is that they are expressed in
terms of the present values of the coupling parameter $\omega$,for
vacuum,radiation and matter dominated epochs  of the Universe.We write gravitational wave perturbations in different epochs of the Universe by cosidering these solutions in terms of time t.
\subsection{Vacuum epoch}
For this epoch  $ \alpha $,$ \beta $,$ n $ are [11]

  $\alpha=\frac{1}{\omega_{0}+2},\beta=-\frac{1}{\omega_{0}+2},n=-(2\omega_{0}+5)$\\
As,$d\eta=\frac{dt}{a}$,we find,$\eta \approx (\frac{2+\omega_{0}}{1+\omega_{0}}) t^{\frac{1+\omega_{0}}{2+\omega_{0}}}$\\
And we take,$(\eta-\overline{\eta_{0}}) \approx (\frac{2+\omega_{0}}{1+\omega_{0}})t^{\frac{1+\omega_{0}}{2+\omega_{0}}}.$\\
Hence,
\[ \mu=\left[ C_{1} \sum_{r=0}^{\infty} A_{r} \left( \frac{\pi}{\lambda}\right)^{m+2r} \left( \frac{2+\omega_{0}}{1+\omega_{0}}\right)^{\frac{1}{2}+(q+1)(m+2r)} t^{(\frac{1+\omega_{0}}{2+\omega_{0}})[\frac{1}{2}+(q+1)(m+2r)]}\right]+\] \\ \[\left[ C_{2} \sum_{r=0}^{\infty} B_{r} \left( \frac{\pi}{\lambda}\right)^{-m+2r} \left( \frac{2+\omega_{0}}{1+\omega_{0}}\right)^{\frac{1}{2}+(q+1)(-m+2r)} t^{(\frac{1+\omega_{0}}{2+\omega_{0}})[\frac{1}{2}+(q+1)(-m+2r)]}\right] \] \\
It is clear from the above expression that,$\omega_{0}\neq -2$ which
is in agreement with the conclusions obtained in our earlier investigation[11].In order to represent $\mu$ a growing modes,the power of  t  should be real and greater than zero.This gives following conditions\\
(i)$\omega_{0}> -1$\\
(ii)$(q+1)(m+2r) >  -\frac{1}{2}$\\
(iii)m should be real.\\
In this case,the order of the Bessel function  is
$m=\pm \frac{1}{2}\sqrt{ 1+ \frac{2\omega_{0}+9}{(\omega_{0}+1)^{2}}}$\\
As m should be real,we get that $\omega_{0}>-\frac{9}{2}$ which is
automatically satisfied if the condition (i) i.e,$\omega_{0} > -1$ is
valid.Thus,we conclude that for gravitational waves to grow
$\omega_{0}$ should take values  greater than -1.

\subsection{Radiation epoch}
For this epoch  $\alpha$,$\beta$ and $n$ are [11]\\
$\alpha=\frac{3}{\omega_{0}+6},\beta=-\frac{3}{\omega_{0}+6},n= -( \frac{2\omega_{0}+3}{3}).$\\
As,$d\eta=\frac{dt}{a}$,we find,$\eta \approx (\frac{6+\omega_{0}}{3+\omega_{0}}) t^{(\frac{3+\omega_{0}}{6+\omega_{0}})}$\\
And we take,$(\eta-\overline{\eta_{0}})\approx (\frac{6+\omega_{0}}{3+\omega_{0}}) t^{(\frac{3+\omega_{0}}{6+\omega_{0}})}.$\\
Hence,\[\mu(k,t)=\left[C_{1} \sum_{r=0}^{\infty} A_{r}\left( \frac{\pi}{\lambda}\right)^{m+2r} \left(\frac{6+\omega_{0}}{3+\omega_{0}}\right)^{\frac{1}{2}+(q+1)(m+2r)} t^{(\frac{3+\omega_{0}}{6+\omega_{0}})[\frac{1}{2}+(q+1)(m+2r)]} \right]+\]\\
\[\left[ C_{2} \sum_{r=0}^{\infty} B_{r} \left( \frac{\pi}{\lambda}\right)^{-m+2r} \left(\frac{6+\omega_{0}}{3+\omega_{0}}\right)^{\frac{1}{2}+(q+1)(-m+2r)} t^{\frac{3+\omega_{0}}{6+\omega_{0}}[\frac{1}{2}+(q+1)(-m+2r)]}\right]\]\\
Thus,for gravitational waves to grow during the radiation epoch the
above solution should satisfy the following conditions,\\
(i) $\omega_{0}>-3$\\ 
(ii) $(q+1)(m+2r) >  -\frac{1}{2}$\\
(iii)m should be real.\\
Using the values of $\alpha$,$\beta$ and $n$ in the expression for m
,we get
$m=\pm\frac{1}{2}\sqrt{1+\frac{6\omega_{0}+9}{(\omega_{0}+3)^{2}}}$.As
m should be real we get $\omega_{0}>-\frac{3}{2}$ which automatically
satisfies constraint(i).Thus,for gravitational waves to grow
$\omega_{0}$ should take values greater than  $-\frac{3}{2}$.\\
$h_{ij}^{k}=\frac{1}{R}  \hspace{.2in}\mu(k,t)\hspace{.2in} \zeta_{ij}(k,t)$\\
$h_{ij}=\int d^{3}k \hspace{.2in} h_{ij}^{k}(k,t)$\\
This represents the perturbation in  Gravitational field  during the radiation epoch.
\subsection{Matter epoch}
During this epoch $\alpha$,$\beta$, and $n$ are [11]\\
$\alpha=\frac{2}{\omega_{0}+4},\beta=- \frac{2}{\omega_{0}+4},n=-(2+\omega_{0})$\\
As,$d\eta=\frac{dt}{a}$,we find,$\eta \approx
(\frac{4+\omega_{0}}{2+\omega_{0}})
t^{(\frac{2+\omega_{0}}{4+\omega_{0}})}$\\
And we take,$(\eta-\overline{\eta_{0}}) \approx
(\frac{4+\omega_{0}}{2+\omega_{0}})
t^{(\frac{2+\omega_{0}}{4+\omega_{0}})}.$\\
Hence,\[
\mu(k,t)=\left[C_{1}\sum_{r=0}^{\infty}A_{r}\left(\frac{\pi}{\lambda}\right)^{m+2r}\left(\frac{4+\omega_{0}}{2+\omega_{0}}\right)^{\frac{1}{2}+(q+1)(m+2r)} 
  t^{(\frac{2+\omega_{0}}{4+\omega_{0}})[\frac{1}{2}+(q+1)(m+2r)]}\right]+\]\\
\[\left[ C_{2} \sum_{r=0}^{\infty} B_{r}
  \left(\frac{\pi}{\lambda}\right)^{-m+2r} \left(
    \frac{4+\omega_{0}}{2+\omega_{0}}\right)^{\frac{1}{2}+(q+1)(-m+2r)} t^{(\frac{2+\omega_{0}}{4+\omega_{0}})[\frac{1}{2}+(q+1)(-m+2r)]}\right]\]\\
It is clear from the above expression that,$\omega_{0}\neq -2$ .Again this
result agrees with our earlier conclusions[11].For growing modes,the
power of t should be real  and greater than zero.This gives following conditions\\
(i)$\omega_{0}>-2$,\\
(ii)$(q+1)(m+2r) >  -\frac{1}{2}$\\
(iii)  m should be real.\\
In this case,the order of the Bessel function,m,is $m=\pm
\frac{1}{2}\sqrt{1+\frac{12+4\omega_{0}}{(2+\omega_{0})^{2}}}.$\\
As m should be real,we get $\omega_{0}>-3$.\\Thus,for perturbation to
grow $\omega_{0} $ should take values greater than -2.
Finally,we get
$h_{ij}^{k}=\frac{1}{R}\hspace{.1in}\mu(k,t)\hspace{.1in}\zeta_{ij}(k,t)$ 
and \\
$h_{ij}=\int d^{3}k\hspace{.1in}h_{ij}^{k}(k,t)$\\
This represents the perturbation in gravitational field  during the matter epochs.
\section{CONCLUSIONS}
We have solved the gravitational wave equation for a general class of
scalar tensor theories of gravity. The solutions are expressed in
terms of the present value of the Brans-Dicke coupling parameter
i.e,$\omega_{0}$.The solutions represent growing modes when the
parameter $\omega_{0}$ takes negative values particularly greater than 
-1.This negative values of the parameter are
required for structure formation,cosmic accleration,superluminal
expansion and at the same time a significant amplification of
perturbations[7,
12].It is also seen that,unlike in GR, there can be amplification of gravitational waves  in a radiation -dominated universe.These new features are consequence of the more general scalar tensor theories of gravity which in special cases reduces to Brans-Dicke thory of gravity and GR.
In a subsequent paper we will use the results derived here  to obtain the spectrum of the relic gravitons in this model.This will enble us to constrain some of the parameters appearing in the scalar -tensor theories. In particular,it could impose a further restriction on the value of the coupling parameter at early times.A small $\omega(\phi)$ is required to have satisfactory nucleation process in some inflationary cosmologies.\\
\centerline{{\bf Acknowldgment}}\\
The authors are grateful to DST,Govt. of India for providing financial
support.The authors also thank Institute of Physics,Bhubaneswar,India
for providing facility of the computer center.

\end{document}